\documentclass[12pt]{iopart}

\usepackage{epsfig}
\usepackage{iopams}  
\begin{document}

\title[Theories of systems with limited information content]{Theories of systems with limited information content}

\author{T Paterek$^{1,2}$, 
B Daki\'{c}$^{2,3}$
and {\v C} Brukner$^{2,3}$}

\address{
$^1$ Centre for Quantum Technologies, National University of Singapore, 3 Science Drive 2, 117542 Singapore \\
$^2$ Institute for Quantum Optics and Quantum Information, Austrian Academy of Sciences, Boltzmanngasse 3, A-1090 Vienna, Austria \\
$^3$ Faculty of Physics, University of Vienna, Boltzmanngasse 5, A-1090 Vienna, Austria}

\ead{tomasz.paterek@nus.edu.sg}

\begin{abstract}
We introduce a hierarchical classification of theories that describe systems with fundamentally limited information content.
This property is introduced in an operational way and gives rise to the existence of mutually complementary measurements,
i.e. a complete knowledge of future outcome in one measurement is at the expense of complete uncertainty in the others. 
This is characteristic feature of the theories and they can be ordered according to the number of mutually complementary measurements
which is also shown to define their computational abilities.
In the theories multipartite states may contain entanglement
and tomography with local measurements is possible.
The classification includes both classical and quantum theory
and also generalized probabilistic theories with higher number of degrees of freedom,
for which operational meaning is given.
We also discuss thought experiments discriminating standard quantum theory from the generalizations.
\end{abstract}

\maketitle

\section{Introduction}

Can one find a class of logically conceivable
physical theories that all share some fundamental features with quantum mechanics? 
For example, in gravitational physics,
general relativity and Brans-Dicke theory~\cite{bransdicke}
belong to a broad class of 
relativistic classical theories of gravitation.
By contrast, it is often assumed that any modification of quantum mechanics 
would produce internally inconsistent theories~\cite{WEINBERG}.

In this paper we identify a class of quantum-like theories
describing systems with limited information content~\cite{zeilinger,BZ2003}. 
This limit does not arise from an observer's ignorance about
the ``true ontic states of reality''~\cite{spekkens} --- which would
be a hidden-variable theory and would have to confront
the theorems of Bell~\cite{bell} and
Kochen-Specker~\cite{kochenspecker} --- but rather is a fundamental limit.
To introduce an operational notion of information content,
we insert the system into a ``black box'', which itself has one of a number of configurations.
After leaving the black box, the system is measured to reveal
some of the properties of the configuration. The ``limited information content of the system'' 
represents the fundamental restriction on how much information about the
configuration can be gained in this measurement.

We first consider a system with an information content of one bit, which we call a two-level system
\footnote{Even if more than two detectors are involved in the measurement of such system,
it can only reveal one bit of information about the configuration in the black box.}.
A measurement outcome can only reveal one bit of information,
i.e. it can distinguish between two equally-sized subsets of possible configurations, 
without any possibility of discriminating between further subsets. 
This gives rise to mutually complementary properties of black box configurations
and the notion of complementary questions, which are questions about these properties.
We study the information gain about these configurations 
which can be revealed using two-level systems described by different theories.
The number of complementary system observables predicted by the theories
limits the number of complementary black-box configurations
which can be accessed.
We use this to identify a hierarchical classification of quantum-like theories.
We show that classical physics --- with no complementary observables --- and quantum physics
--- with three complementary observables for a qubit --- are just two
examples of theories within this hierarchy and present examples of other theories.
A theory on a particular level of the hierarchy contains all lower-level theories,
just as theory of quantum bits contains theory of classical bits.

We investigate the computational capabilities of the new theories
in a manner similar to the work on no-signaling theories \cite{MAG2006,NONLOCAL_COMP,BARRETT,BARNUM1,BARNUM2,BARNUM3,VSW2009}
and show that computational capabilities increase with the level of the theory in the hierarchy.
We then consider composite systems, and demonstrate
existence of complementary properties of many black boxes
which cannot be accessed with (product of) independent subsystems,
leading to necessity of entanglement in the corresponding theories.
We also show that the number of parameters obtained from complementary
measurements on a composite system consisting of many two-level systems 
agrees with the number of parameters obtained from correlations 
between complementary local measurements.
This fact is a remarkable coincidence since a priori there is nothing in the definition of the hierarchy that hints at it.
Finally, we present thought experiments aimed at distinguishing
standard quantum theory from the generalized theories.

Other attempts have previously been made to introduce a hierarchy of models that includes both classical and quantum theory.
The generalized models exploit different sum-rules for probabilities~\cite{SORKIN}
or explore physical systems described by a number of parameters (sometimes also called ``degrees of freedom'')
different than in quantum mechanics~\cite{HIERARCHY_WOOTTERS,HIERARCHY_HARDY,HIERARCHY_ZYCZKOWSKI,HIERARCHY_FUCHS,DB}.
Our approach is related to the later in that we consider two-level systems
with additional degrees of freedom. 
We show that the principle of limited information content
together with an assumption that a system can reveal any of the complementary
properties of black box configurations allows only specific values for the number of these degrees.
The same number is derived by Wootters \cite{HIERARCHY_WOOTTERS} and Hardy \cite{HIERARCHY_HARDY}
using parameter counting argument for composite systems.
Here, however, it follows already for a single system.

It should be noted that our aim here is not to derive the structure of quantum theory
but rather to show alternative models whose parameters
also have operational meaning.
It is interesting to ask which axioms of standard quantum theory such models defy.
Compared with Hardy's axiomatization \cite{HIERARCHY_HARDY},
our models for a single two-level system involve more degrees of freedom than a qubit
and therefore include also those theories which Hardy excluded by the simplicity axiom
(the simplicity axiom states that one should take the minimal number of degrees of freedom in agreement with other axioms).
The probability axiom (in all experiments on a sufficiently big ensemble of systems prepared in the same way, the relative frequencies of measurement outcomes tend to the same values)
is fulfilled in our models.
The continuity axiom (there exists a continuous reversible transformation on a system between any two pure states of that system),
is fulfilled by the presented models of a single system.
For multiple two-level systems, assumption of limited information content
together with requirement that systems reveal any of complementary properties
implies Hardy's axiom about composite systems (local tomography is possible).
It states that both the number of levels of a composite system, $N$,
and the number of parameters describing its unnormalized states, $K$, are products
of respective numbers for individual subsystems, i.e. $N = N_A N_B$ and $K = K_A K_B$.
It was proved that Hardy's simplicity axiom is redundant \cite{DB},
i.e. that only classical and quantum theories are in agreement with all other axioms.
This implies for the multipartite theories studied here that they have to defy Hardy's subspace axiom
(it states that a $n$-level subsystem of a higher-level system behaves like a system with $n$ levels).
This is a consequence of the fact that continuity is fulfilled by the presented models for a single system
and therefore the subspace axiom implies continuity for many systems because
any two states of a composite system are connected by a continuous transformation, introduced in a single particle case.
As we already noted, this would constrain the possible theories to classical and quantum only due to the results of Ref. \cite{DB}.

\section{Limited information content}

\begin{figure}
\begin{center}
\includegraphics[scale=0.4]{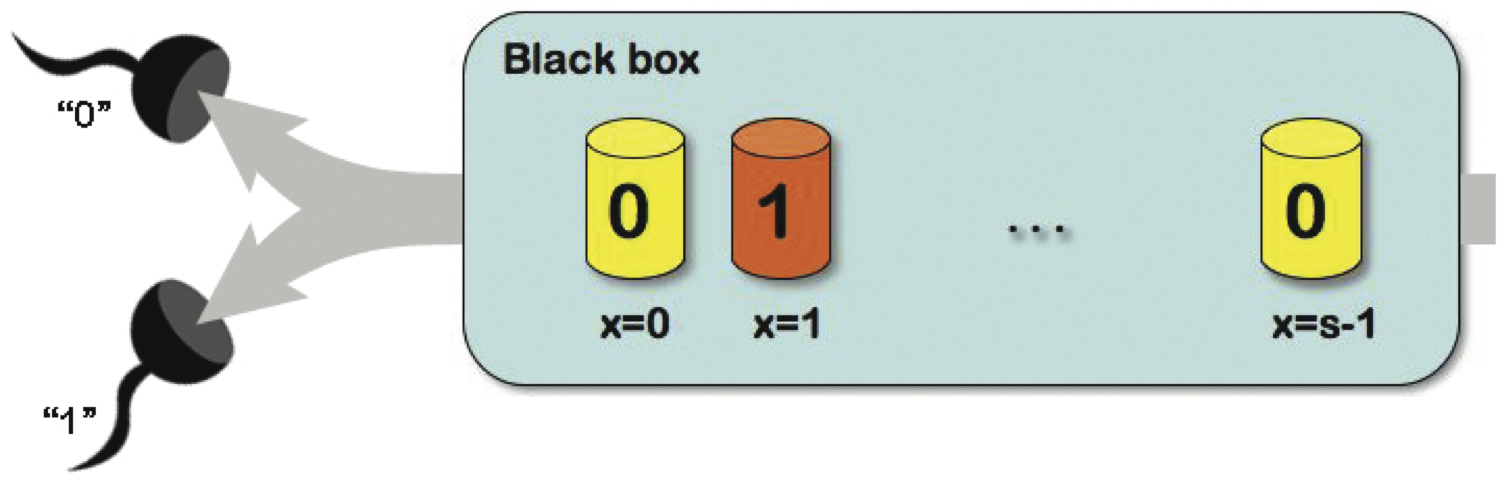}
\end{center}
\caption{The configuration of items inside the black box is a physical realization of a function $y =
f(x)$. The value of $x$ is encoded in the position inside the box,
whereas the value of $y$ is encoded by putting a yellow ($y=0$) or
orange ($y=1$) item at position $x$. A physical system enters the black box from the right,
undergoes function-dependent transformations and is finally
measured after leaving the box.}
\label{FIG_BB}
\end{figure}

Consider the black box
illustrated in Fig.~\ref{FIG_BB}. 
A Boolean function of a single $s$-valued argument, 
$y = f(x)$, with $x = 0, ..., s-1$ and $y = 0,1$,
is realized physically by putting one of two different (classical)
objects in each of $s$ different positions inside the box. 
As a result, there are $2^s$ different functions $f(x)$
and as many distinguishable configurations of the black box. 
If all the configurations have the
same probability of occurring, $s$ bits of information are
necessary to identify a given function. 
A physical system with information content below $s$ bits cannot therefore distinguish an individual function,
but only groups of functions with certain properties.

For example, consider a black box with two positions inside
which is probed by a single two-level system.
The possible box configurations represent four Boolean functions
of the position variable $x = 0,1$, which can be indexed by $j = 2^1 f(0) + 2^0 f(1)$. 
The readout step reveals one bit of information, splitting the four functions into two equally-sized sets. 
In this case, one finds three possible splits which can be illustrated by the three rows of the
following tables (symbol $\oplus$ denotes addition modulo two):
\begin{equation}
\begin{tabular}[b]{cc|cc}
\hline  \hline
$0$ & $1$ & $2$ & $3$
\\ \hline
$0$ & $2$ & $1$ & $3$
\\ \hline
$0$ & $3$ & $1$ & $2$
\\ \hline \hline
\end{tabular}
\quad  \quad
\begin{tabular}[b]{cc|cc}
\multicolumn{2}{c|}{$a=0$} & \multicolumn{2}{c}{$a=1$} \\
\hline  \hline
$00$ & $01$ & $10$ & $11$
\\ \hline
$00$ & $10$ & $01$ & $11$
\\ \hline
$00$ & $11$ & $01$ & $10$
\\ \hline \hline
\end{tabular}
\quad  \quad
\begin{tabular}[b]{c}
\hline  \hline
$f(0) = a?$
\\ \hline
$f(1) = a?$
\\ \hline
$f(0) \oplus f(1) = a?$
\\ \hline \hline
\end{tabular}
\label{2DESIGN}
\end{equation}
The table on the left-hand side shows the index $j$
\footnote{An equivalent table was introduced by Spekkens within a different
interpretational approach~\cite{spekkens}.
There, an individual quantum system is assumed to be in an ontic state,
while here only the (classical) black box is in a well-defined ``ontic'' state.},
and the middle table shows the functional values ordered as pairs $f(0)\,f(1)$. 
The table on the right-hand side gives the three complementary questions about the properties of the functions.
They are answered by the functions  in the left and right column of the middle table
(left column $\to$ answer $0$, right column $\to$ answer $1$).
We shall refer to such tables as the complementarity tables \cite{MUBOLS}.

The black box forms a bridge between the abstract
mathematical construction of complementarity tables and the physical world. 
The physical system 
can be used to probe the box configuration
by subjecting it to configuration-dependent transformations.
An appropriate measurement
can then be used to identify the subset to which the configuration belongs. 
Two-level systems described by different physical theories
allow one to answer different numbers of complementary questions.

In the simplest case, $s=1$, the black box contains only one position.
It is convenient to think of
the value $f(0)=0$ as an empty position and $f(0)=1$ as an occupied position.
This configuration can be revealed by a classical bit,
which by definition can only either be flipped or left untouched.
If its state is flipped only when the object is present,
then knowing the initial and final states
of the bit completely determines the box configuration, $f(0)$.
This is possible because the box stores only one bit.

The next case, with two positions inside the black box, is qualitatively different
because complementary questions now arise.
A classical bit can no longer be used to answer any one of them. 
This can, however, be achieved using a quantum bit.

A quantum bit can be entirely expressed
in terms of real vectors in three dimensions.
The set of pure quantum states forms  
a unit Bloch sphere, with orthogonal axes
representing the eigenstates of complementary observables.
The set of operations on a qubit is no longer restricted only to bit flips, 
but includes any rotation.
Consider the following interaction between the system and the black box.
For $f(x) = 0$ (position $x$ is empty), the qubit state is left untouched. 
If $f(x)=1$ (occupied), the $\sigma_x$ or $\sigma_z$ Pauli rotation is applied
to the qubit state for $x=0$ or $1$, respectively.
The qubit propagates through the black box from right to left,
giving a total transformation of $\sigma_x^{f(0)} \sigma_z^{f(1)}$. 
In Bloch coordinates, these rotations are represented by diagonal matrices, 
$\sigma_x \to \mathrm{diag}[1,-1,-1]$ and $\sigma_z \to \mathrm{diag}[-1,-1,1]$.
Thus, the interaction of the black box with the system 
is represented by the diagonal matrix
\begin{equation}
\mathrm{diag}[(-1)^{f(1)},(-1)^{f(0)+f(1)}, (-1)^{f(0)}].
\end{equation}
The quantum probability to observe an outcome associated with the state $\vec m$, 
given a system prepared in state $\vec n$, is $P(\vec m | \vec n) = \frac{1}{2}(1 + \vec n
\cdot \vec m)$, where the dot denotes a scalar product in $\mathbb{R}^3$. 
Therefore, if the $| z \pm \rangle$ states are used as inputs, 
the measurement in this basis after the interaction reveals the value of $f(0)$.
Similarly, using $| x \pm \rangle$ or $| y \pm \rangle$ as inputs,
and measuring in these bases,
reveals the value of $f(1)$ and $f(0) \oplus f(1)$, respectively.
Thus, each of the complementary questions can be answered using
the eigenstates of the complementary quantum observables.

\section{Generalized theories}

We next investigate a black box containing three positions, $x=0,1,2$.
The resulting complementarity table has \emph{seven} rows:
\begin{equation}
\begin{tabular}{cccc|cccc}
\hline  \hline
$0$ & $1$ & $2$ & $3$ & $4$ & $5$ & $6$ & $7$
\\ \hline
$0$ & $1$ & $4$ & $5$ & $2$ & $3$ & $6$ & $7$
\\ \hline
$0$ & $1$ & $6$ & $7$ & $2$ & $3$ & $4$ & $5$
\\ \hline
$0$ & $2$ & $4$ & $6$ & $1$ & $3$ & $5$ & $7$
\\ \hline
$0$ & $2$ & $5$ & $7$ & $1$ & $3$ & $4$ & $6$
\\ \hline
$0$ & $3$ & $4$ & $7$ & $1$ & $2$ & $5$ & $6$
\\ \hline
$0$ & $3$ & $5$ & $6$ & $1$ & $2$ & $4$ & $7$
\\ \hline \hline
\end{tabular}
%\quad \quad
%\begin{tabular}{cccc|cccc}
%\hline  \hline
%$000$ & $001$ & $010$ & $011$ & $100$ & $101$ & $110$ & $111$
%\\ \hline
%$000$ & $001$ & $100$ & $101$ & $010$ & $011$ & $110$ & $111$
%\\ \hline
%$000$ & $010$ & $100$ & $110$ & $001$ & $011$ & $101$ & $111$
%\\ \hline
%$000$ & $001$ & $110$ & $111$ & $010$ & $011$ & $100$ & $101$
%\\ \hline
%$000$ & $010$ & $101$ & $111$ & $001$ & $011$ & $100$ & $110$
%\\ \hline
%$000$ & $011$ & $100$ & $111$ & $001$ & $010$ & $101$ & $110$
%\\ \hline
%$000$ & $011$ & $101$ & $110$ & $001$ & $010$ & $100$ & $111$
%\\ \hline \hline
%\end{tabular}
\quad \quad
\begin{tabular}{c}
\hline  \hline
$f(0)=?$
\\ \hline
$f(1)=?$
\\ \hline
$f(2)=?$
\\ \hline
$f(0) \oplus f(1)=?$
\\ \hline
$f(0) \oplus f(2)=?$
\\ \hline
$f(1) \oplus f(2)=?$
\\ \hline
$f(0) \oplus f(1) \oplus f(2)=?$
\\ \hline \hline
\end{tabular}
\label{POSTQUANTUM}
\end{equation}
The table on the left-hand side presents the values of $j=2^2 f(0)+2^1 f(1)+2^0 f(2)$.
Given one bit of information that
answers any single complementary question in the right-hand-side table, 
no information can be obtained about an answer to any of the other questions,
i.e. the seven questions are logically independent \cite{INDEPENDENCE}.

In analogy to the previous cases, one can ask what ``physical theory'' for the system
is required to answer any one of the complementary questions contained in table (\ref{POSTQUANTUM}).
Such a theory must contain features of complementarity,
and we now generalize the Bloch representation of a quantum bit
to produce a quantum-like theory
related to the black box with three internal positions.
Since there are seven complementary questions, 
there must be seven complementary measurements for the system
and we assume its pure physical states 
are represented by vectors on a sphere
in seven dimensions (state space postulate).
Given a system prepared in a state $\vec n$,
the probability to observe an outcome associated with the state $\vec m$,
is chosen as $P(\vec m | \vec n) = \frac{1}{2}(1 + \vec n \cdot \vec m)$, where the dot
now denotes a scalar product in $\mathbb{R}^7$ (probability rule).
To fulfill the physical requirement that immediate repetition of the same measurement
should have the same outcome, the state $\vec n$ 
is updated in the measurement to $+ \vec m$ or $- \vec m$, depending on the result (collapse postulate).
The physical transformations, including temporal evolution, are represented in this theory by rotations belonging to SO$(7)$.
They preserve distinguishability between any two states
as measured by the scalar product, 
and are continuously connected with the identity, i.e., no transformation.

The model just described allows us to answer any complementary question from table (\ref{POSTQUANTUM}).
The black box transformation can be chosen to be a product
$R_0^{f(0)} R_1^{f(1)} R_2^{f(2)}$ of rotations
\begin{eqnarray}
R_0 & \to & \mathrm{diag}[-1,1,1,-1,-1,1,-1], \nonumber \\
R_1 & \to & \mathrm{diag}[1,-1,1,-1,1,-1,-1], \nonumber \\
R_2 & \to & \mathrm{diag}[1,1,-1,1,-1,-1,-1].
\end{eqnarray}
This product is a diagonal matrix with seven entries:
$(-1)^{f(0)}$, $(-1)^{f(1)}$, $(-1)^{f(2)}$, $(-1)^{f(0)+f(1)}$,
$(-1)^{f(0)+f(2)}$, $(-1)^{f(1)+f(2)}$, $(-1)^{f(0)+f(1)+f(2)}$,
where the powers are specified by the complementary questions.
Therefore, to answer a complementary question one propagates through the black box
system prepared  in a state related to the corresponding complementary measurement
and finally performs this measurement.

In the general case of a black box with $s$ internal positions, 
one finds ${s \choose 1} + {s \choose 2} + ... + {s \choose s} = 2^s{ - }1$ complementary questions. 
There are ${s \choose 1}$ questions about the value of
$f(x)$, ${s \choose 2}$ questions about different sums of $f(x)
\oplus f(x')$ with $x \ne x'$, and so forth. 
A physical theory of a two-level
system can be constructed with $2^s{ - }1$ complementary measurements
using the approach described above.
Since $s$ can be arbitrarily large,
there are complementarity tables with arbitrarily many rows,
and correspondingly many different theories for a two-level system.

Importantly, the derived number of independent parameters
which completely specify the state in a generalized theory,
i.e. $2^s-1$, is the same as the one
following from the parameter counting argument for composite systems \cite{HIERARCHY_WOOTTERS,HIERARCHY_HARDY}.
Here, however, it follows already for a single system:
from the operational definition (via black box)
of the limited information content
and the assumption that a system can answer any of the complementary questions.

In all cases, the quantum-like models we have introduced
possess rotationally invariant state spaces.
There is therefore no preferred choice of a set of $2^s-1$ complementary directions or any preferred state.
One may expect information contained in all pure states $\vec n$ to be the same
and independent of the choice of a complete set of complementary measurements.
We ask how to quantify information gain in a single measurement
$I(p_{+j},p_{-j})$, with $p_{\pm j} = \frac{1}{2}(1 \pm \vec n \cdot \vec m_j)$
being probabilities for $\pm 1$ results in measurement $\vec m_j$,
such that this expectation is fulfilled.
Assuming after Ref. \cite{MEASURE} that information content of state $\vec n$
is the sum of information gained in all complementary measurements
$I(\vec n) = \sum_{j =1}^{2^s - 1} I(p_{+j},p_{-j})$
the argument of Ref. \cite{BZ_UNIQUE} shows that
in the set of information measures based on $\alpha$-entropy,
i.e. if one takes $I(p_{+j},p_{-j}) = 1 - k \frac{1- p_{+j}^\alpha - p_{-j}^\alpha}{\alpha -1}$ with a constant $k$ and real parameter $\alpha$,
only for the quadratic measure, with $\alpha = 2$, the information content $I(\vec n)$
is constant and invariant under a \emph{continuous} change between different complete sets of mutually complementary directions.
Fixing $k=2$ sets the units such that we have $I(n_j) = n_j^2$, 
where $n_j = \vec n \cdot \vec m_j$ and since the directions of complementary measurements
are orthogonal one finds $I(\vec n) = |\vec n|^2$,
which immediately generalizes the measure of Ref. \cite{MEASURE}.
This measure captures intuitive expectation that overall information  contained in a pure state (revealed in the complete set of complementary measurements) is again one bit.

\section{Computational abilities of generalized theories}

The theories with different
number of complementary measurements have different computational
abilities. Consider the problem of determining properties of a 
function with a single query of the black box. 
As an example, think about table (\ref{2DESIGN}). 
A qubit propagating through
the black box is able to reveal the value of any of
$f(0)$, $f(1)$ or $f(0) \oplus f(1)$ by making the appropriate choice of input state and measurement~\cite{DEUTSCH}.
Classically this is impossible.
A classical bit can in principle reveal only one of the three properties
because each of the items inside the black box can either keep the bit value or flip it.
For example, if the classical bit is flipped after leaving the box,
then we know that one of the internal positions is occupied,
but it is impossible to determine which one no matter what initial state is used.

Likewise, table (\ref{POSTQUANTUM}) illustrates the limitations of quantum computing. 
A single two-level system with seven complementary observables 
can encode an answer to any one of the seven complementary questions.
By contrast, it is only possible to answer
at most three of the questions using one qubit.
A qubit can be embedded into all generalized theories,
just as classical bit is embedded into quantum theory.
A sphere in $2^s-1$ dimensions, for $s>2$, always contains as subspace
a two-sphere of pure states of a quantum bit,
and rotations on a two-sphere are a subset of all rotations on higher-dimensional spheres.
The rotations of two-sphere, when applied in arbitrary order,
never evolve the system outside the two-sphere.
Therefore, even if the qubit interacts with more than two items in a black box,
it can never answer more than three complementary questions.
All generalized theories with more complementary observables
are computationally more powerful than both classical and quantum physics.

\section{Many systems}

The presentation so far has been limited to a single system.
We operationally define the information content of $N$ systems
as a maximal possible information gain
about the internal configuration of $N$ black boxes, each for a single system.
Therefore, the information content of $N$ two-level systems
is limited to $N$ bits \cite{zeilinger}.
We show that the number of independent real parameters obtained
from (joint) complementary measurements,
answering the questions about the complementary properties of 
$N$ Boolean functions encoded in the black boxes, 
is the same as the number of parameters obtained from correlations
between local complementary measurements.

To simplify the presentation we start with two quantum systems as an illustration of ideas and techniques,
and next give general results \footnote{For the simplest non-classical and non-quantum example, $N=2$ and $s=3$, the complementarity table has $21$ rows and it is cumbersome to present it explicitly.}.
The quantum case corresponds to $s=2$.
For two qubits we have two black boxes, 
each of which encodes one of four Boolean functions, see (\ref{2DESIGN}),
and therefore there are in total $2^{Ns} = 16$ combinations of pairs of functions in two black boxes.
Accordingly, every row of the complementarity table contains $16$ items.
Since in this case the final measurement reveals two bits of information,
the table has $2^N = 4$ columns.
Complementary properties of two Boolean functions are defined such that full knowledge of one property
precludes any knowledge about the other property.
They correspond to the rows of the table in which items
from a fixed column of one row (full knowledge)
are evenly distributed among all columns of any other row (no knowledge).
For example, for two qubits we have:
\begin{equation}
\begin{tabular}{cccc|cccc|cccc|cccc}
\multicolumn{4}{c|}{$a_1=0 \quad a_2 = 0$} & \multicolumn{4}{c|}{$a_1=0 \quad a_2 = 1$} & \multicolumn{4}{c|}{$a_1=1 \quad a_2 = 0$} & \multicolumn{4}{c}{$a_1=1 \quad a_2 = 1$}\\
\hline  \hline
$00$ & $01$ & $10$ & $11$
& $02$& $03$& $12$& $13$
& $20$& $21$& $30$& $31$
& $22$& $23$& $32$& $33$
\\ \hline 
$00$ & $02$ & $20$ & $22$
& $01$& $03$& $21$& $23$
& $10$& $12$& $30$& $32$
& $11$& $13$& $31$& $33$
\\ \hline 
$00$ & $03$ & $30$ & $33$
& $01$& $02$& $31$& $32$
& $10$& $13$& $20$& $23$
& $11$& $12$& $21$& $22$
\\ \hline 
$00$ & $12$ & $23$ & $31$
& $02$& $10$& $21$& $33$
& $01$& $13$& $22$& $30$
& $03$& $11$& $20$& $32$
\\ \hline 
$00$ & $13$ & $21$ & $32$
& $01$& $12$& $20$& $33$
& $02$& $11$& $23$& $30$
& $03$& $10$& $22$& $31$
\\ \hline \hline
\end{tabular}
\label{4DESIGN}
\end{equation}
where each item is a pair of numbers $j_1 \, j_2$
describing functions in the first and second black box respectively,
i.e. $j_1 = 2 f_1(0) + f_1 (1)$ and $j_2 = 2 f_2(0) + f_2 (1)$.
The complementary properties in this case are the following:
(i) the first row corresponds to two binary questions, whether $f_1(0) = a_1$ and $f_2(0) = a_2$,
(ii) the second row corresponds to asking whether $f_1(1) = a_1$ and $f_2(1) = a_2$,
(iii) the third row is the ``parity question'', whether $f_1(0) \oplus f_1(1) = a_1$ and $f_2(0) \oplus f_2(1) = a_2$,
(iv) the forth row coincides with asking whether $f_1(0) \oplus f_2(1) = a_1$ and $f_1(0) \oplus f_1(1) \oplus f_2(0) = a_2$,
(v) the last row leads to asking if $f_1(1) \oplus f_2(0) = a_1$ and $f_1(0) \oplus f_1(1) \oplus f_2(1) = a_2$.
The answers to these questions are in a form of two bit values $a_1 \, a_2$
and the columns of the table from left to right correspond to the answers $00$, $01$, $10$ and $11$.
Such complementarity tables are well-known in a mathematical theory of combinatorial designs.
In the quantum case of $s=2$ they are so-called net designs,
and the maximal number of their rows gives the number
of complementary quantum measurements \cite{MUBOLS}.
In a general case of arbitrary $s$, 
the complementarity table describing complementary properties of $N$ Boolean functions of an $s$-valued argument
has $2^{Ns}$ items in every row and $2^N$ columns.
Such complementarity tables, with $s>2$, are known as the generalized net designs (affine $1$-designs)
and the maximal number of their rows is given by the Bose-Bush bound \footnote{Pages 219-220 of Ref. \cite{CRC}.
In their notation, $\lambda = 2^{N(s-2)}$, $v=n=2^N$
and $k$ is the number of rows in the complementarity table.}:
\begin{equation}
r_s(N) = \frac{2^{Ns}-1}{2^N-1}.
\end{equation}
Each of the $r_s(N)$ mutually complementary (joint)
measurements gives $2^N - 1$ independent real parameters
(due to normalization) and therefore all the complementary measurements
give altogether $r_s(N)(2^N-1) = 2^{Ns}-1$ independent real parameters.

The same number is found via ``tomography with local measurements'' \cite{HIERARCHY_WOOTTERS, HIERARCHY_HARDY},
in which case we are looking into correlations
between the outcomes of all combinations of complementary local measurements (on every subsystem).
Each single system is described by $2^s-1$ real parameters.
Additionally, one measures correlations between $2,3,...,N$ subsystems
(if none of the subsystems is measured, no information is gained).
This gives $(2^s-1+1)^N-1 = 2^{Ns}-1$ independent real parameters.
Thus, we have shown that the number of  parameters obtained from joint and local measurements coincide.
We see it as an argument that this number of parameters should completely specify a state of the system.
Under this assumption, the models considered
possess an intuitive feature that a physical state
is equally well described by joint and individual measurements.
These are then just two different ways of accessing the same information about the system.
The equality of the number of parameters obtained by joint and local measurements
also means that the models satisfy Hardy's axiom about composite systems:
the number of levels of the whole system is a product of number of levels of subsystems
and the number of parameters specifying the unnormalized joint state is also a product
of the number of such parameters for the subsystems \cite{HIERARCHY_HARDY}.

The complementary questions related to table (\ref{4DESIGN}) and similar tables for many two-level systems
in the generalized theories reveal that the theories involve entanglement.
One can recognize the first three questions of table (\ref{4DESIGN}) are just combinations of
complementary questions for single systems, see (\ref{2DESIGN}).
They are asked independently on every subsystem,
i.e. the questions with the answer $a_1$ involve only function $f_1(x)$ and
the questions with answer $a_2$ involve only function $f_2(x)$.
With them, all the complementary questions for single subsystems are already exhausted.
The same argument applies to any complementarity table of higher level theories.
Since for any such table related to many black boxes
the maximal number of rows is greater than the number of rows
of the table for a single system, there are complementary questions involving relational properties of
functions encoded in different black boxes, such as e.g. the question of the value of $f_1(0) \oplus f_2(1)$ and $f_1(0) \oplus f_1(1) \oplus f_2(0)$.
These questions cannot be answered by systems in a product state
and we conclude that entanglement must be present in such models.

\section{Experimental consequences}

We give two experimental consequences of the generalized theories
that differ from predictions of standard quantum theory of a single two-level system.
Note that if the experimenter has access to generalized states, evolutions and measurements
it is clear that standard quantum theory could be refuted.
It is more realistic however to study if the other models can be identified
by looking only at the data gathered in quantum measurements.
A reason for this is that we now only know how to build apparatuses
corresponding to quantum measurements.
Furthermore, one can imagine that there is in Nature a source emitting states of generalized theories
whereas we are still restricted to quantum measuring devices.
Therefore, we make here an assumption that experimentalists have access only
to measurements allowed by standard quantum mechanics (on the Bloch sphere)
whereas states and evolutions obey generalized theories (on higher dimensional spheres).

The first consequence is a change of purity of an evolving closed system.
When the system represented by a vector in a higher dimensional Bloch sphere evolves in time, 
the projected vector onto the standard two-sphere will in general change its length indicating ``decoherence'' and ``recoherence''
in the effective quantum state description. 
These effects would be present even when the system is closed and can be considered as isolated from environment
according to all means of standard quantum theory.

Second, we present a gedanken experiment which tests
a dimension of the sphere of states.
Consider a scenario in which there are grounds to assume a source
prepares random states
from the entire higher dimensional Bloch ball (also mixed states)
in such a way that the mean value of measurement along some 
$\vec x$ axis can be found for every random state.
For example, the source is slowly randomly evolving
such that within a short time interval the states emitted are basically the same,
but if one waits a longer time and then measures again,
the observed state will be unrelated to the previously observed one.
The frequency with which a mean $\langle \vec x \rangle$ occurs, $f(\langle \vec x \rangle)$, is proportional to the number
of states giving rise to this particular value of $\langle \vec x \rangle$,
which is related to the projection of the state vector on the $\vec x$ axis.
Since the higher the dimension of the sphere
the more states have the mean $\langle \vec x \rangle$ close to zero,
the shape of $f(\langle \vec x \rangle)$ reveals the dimension.
We now develop this idea quantitatively.

\begin{figure}
\begin{center}
$\qquad \qquad$
\includegraphics[scale=0.5]{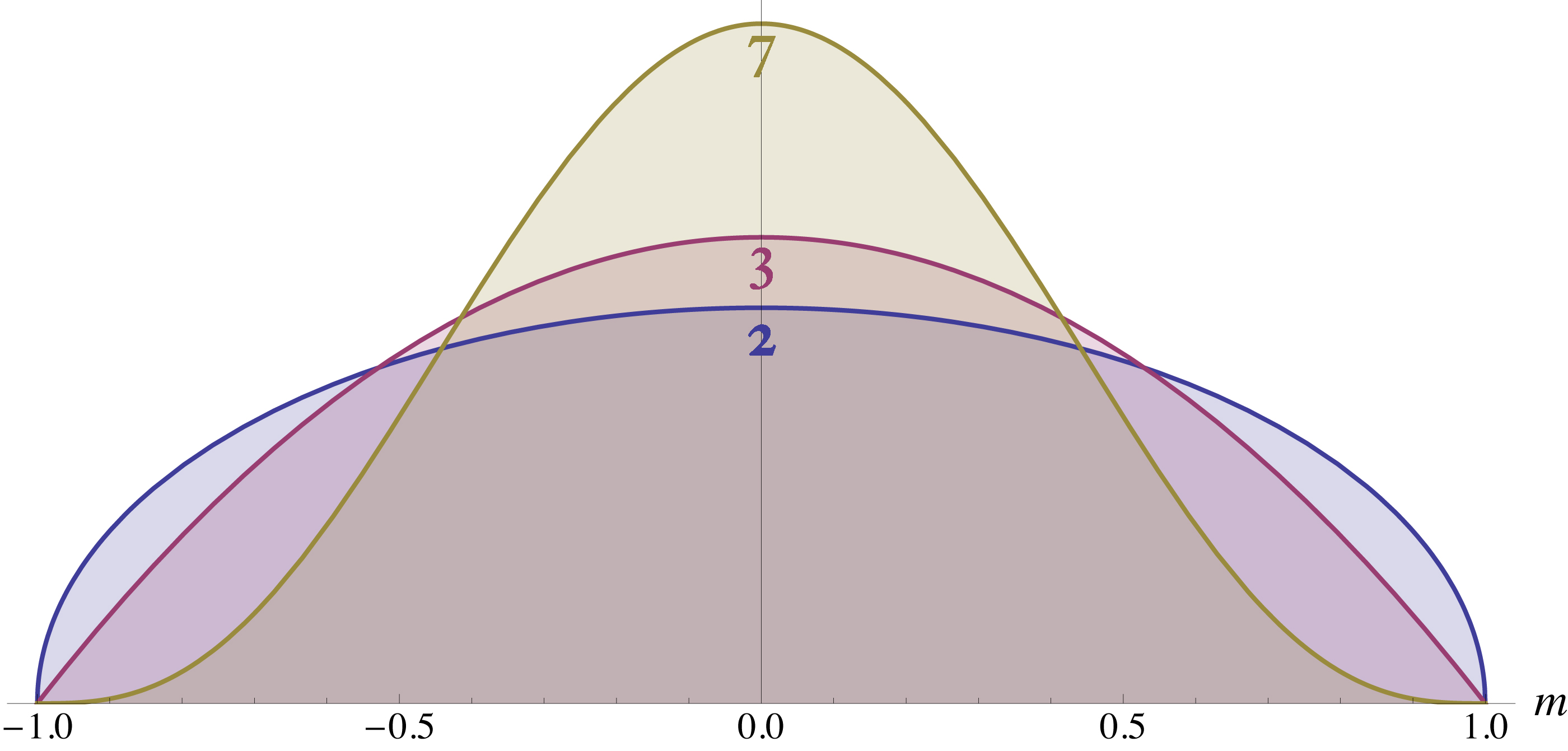}
\end{center}
\caption{Detecting dimension of state space with a random sampler.
Assuming that states are represented by vectors within higher-dimensional sphere,
sampling them randomly in such a way that for each state
the average value, $m$, along some direction $\vec x$ can be measured,
provides a way to find the dimension.
The dimension can be read from the histogram of $m$.
The plot shows the histogram for three dimensions, $D=2$, $3$, and $7$.
Generally, after measuring the frequency of the average values
one finds the dimension from the fit of the curve (\ref{FIND_D}).}
\label{FIG_SAMPLER}
\end{figure}

To make an illustration, we first describe how to distinguish between a theory
in which all the states are within a disk (real quantum theory) and standard (complex) quantum theory having a three-dimensional ball of allowed states.
If the state space is a disc, a random state is distributed with probability density
$dp(x,y) = dx dy / \pi R^2$, where $R$ is the radius of the disc.
The frequency of observation of the average value $m$
in a measurement of $\vec x$ is related to the length
of the chord perpendicular to the $x$ axis which crosses the axis at point $m$,
$F_2(m) = 2 \int_0^{\sqrt{R^2 - m^2}} \frac{dy}{\pi R^2} = \frac{2 \sqrt{R^2 - m^2}}{\pi R^2}$.
If the state space is a ball, a random state is distributed with probability density
$dp(x,y,z) = dxdydz / \frac{4}{3} \pi R^3$,
and the frequency of observation of the average value $m$
is now related to the area of the disc orthogonal to $x$ axis
which crosses the axis at point $m$,
$F_3(m) = \frac{\pi r^2}{\frac{4}{3} \pi R^3}$,
where $r=\sqrt{R^2 - m^2}$ is the radius of the disc.
In general, for a state space which is a sphere in $D$ dimensions,
a random state is distributed according to probability density
$dp(x_1,\dots,x_D) = d x_1 \dots dx_D/V_D(R)$,
where $V_D(R) = \frac{\pi^{D/2} R^D}{\Gamma(D/2 + 1)}$ is the volume of the sphere
embedded in $D$ dimensions
and $\Gamma(x)$ is the gamma function.
The frequency of the average value $m$ is given by the ratio of volumes
$F_D(m) = \frac{V_{D-1}(r)}{V_D(R)}$ with $r = \sqrt{R^2 - m^2}$.
Putting in the explicit formulae for the volumes gives
\begin{equation}
F_D(m) = \frac{1}{\beta(\frac{D}{2} + \frac{1}{2},\frac{1}{2})} \frac{(R^2 - m^2)^{\frac{D-1}{2}}}{R^D},
\label{FIND_D}
\end{equation}
where $\beta(x,y) = \frac{\Gamma(x) \Gamma(y)}{\Gamma(x+y)}$ is the Euler beta function
and we used $\Gamma(1/2) = \sqrt{\pi}$.
Fig. \ref{FIG_SAMPLER} shows $F_D(m)$ for various $D$ and $R = 1$.
Note that in principle $D$ does not even have to be an integer.

If one measures not along a single direction,
but along $d$ orthogonal directions,
the immediate generalization of the frequency formula (\ref{FIND_D}) reads
$F_D(m_1,\dots,m_d) = \frac{V_{D-d}(r)}{V_D(R)}$ with $r = \sqrt{R^2 - m_1^2 - \dots - m_d^2}$.
This can be useful if a random state is not sampled from spherically symmetric space,
providing a way to distinguish even more general models than those studied here.
As an illustration, consider first a single $\vec x$ measurement
and states sampled from a disc. We already know the distribution of $m$ is $F_2(m) = \frac{2 \sqrt{R^2 - m^2}}{\pi R^2}$.
The same distribution is obtained for the state space
which is a half disc cut at the $x$ axis,
because both the probability density for state distribution
and the probability for the mean value equal to $m$
are half those for the disc space and their contributions cancel out in the fraction.
Clearly, measurement along $x$ and $y$ could distinguish these two cases.

\section{Conclusions}

In conclusion,  we have introduced a hierarchy of theories describing systems
with limited information content,
which contains classical and quantum mechanics as special cases.
The order parameter of the hierarchy is the number
of complementary questions about the properties of Boolean functions
the systems described by the theory can experimentally answer.
Typical quantum features such as irreducible randomness and complementarity inevitably occur in the theories.
We consider a physical system
able to encode the answer to any one of the complementary questions,
and assume there is a measuring device which can reveal this information.
While the appropriate measurement will reveal the answer
to the selected question, the complementary measurements must reveal no information whatsoever
--- the readout has to give a completely random answer \cite{INDEPENDENCE}.
Further, since the information content of the system is fundamentally limited to one bit,
no underlying hidden structure (in the form of hidden variables) is possible,
and the results are irreducibly random.
As a final remark, we note that 
we gave examples of generalized theories which
share some essential features with quantum mechanics
but nevertheless differ from it.
Intriguingly, this perhaps suggests that either Nature admits
additional conceptual ingredients that single out quantum theory from the more general class \cite{DB}
or the alternatives are also realized in some domain that is still beyond our observations.

\ack

We are grateful to N. K. Langford
for his careful reading of the manuscript and important remarks.
We also acknowledge discussions with J. Kofler and A. Zeilinger.
This work is supported by the Austrian Science Foundation within SFB,
FWF Project No. P19570-N16 and Doctoral Program CoQuS, 
Foundational Questions Institute (FQXi),
the European Commission, Project QAP (No. 015846),
and the National Research Foundation and Ministry of Education in Singapore.

\end{document}